\documentclass{sig-alternate-05-2015}

\pdfoutput=1

\usepackage{todonotes}
\usepackage{graphicx}
\usepackage{listings}
\usepackage{siunitx}
\usepackage{nicefrac}
\usepackage{float}
\usepackage[T1]{fontenc}
\usepackage[utf8]{inputenc}
\usepackage{algorithm}
\usepackage{algpseudocode}
\usepackage{xfrac}

\algblockx{StartTEE}{EndTEE}[1][enclave]{\textbf{inside TEE} $\textsc{#1}$}{\textbf{end inside TEE}}
\algblockx{StartNoTEE}{EndNoTEE}{\textbf{outside TEE}}{\textbf{end outside TEE}}
\algblockx{StartLoop}{EndLoop}{\textbf{loop}}{\textbf{end loop}}
\algblockx{StartOn}{EndOn}[1]{\textbf{on} #1}{\textbf{end on}}
\algnewcommand\Receive{\textbf{receive} } % Space is needed.
\algnewcommand\Send{\textbf{send} } % Space is needed.
\algnewcommand\Assert{\textbf{assert} } % Space is needed.
\algnewcommand\Yield{\textbf{yield} } % Space is needed.
\algnewcommand\Commentx[1]{\textcolor{gray}{//~#1}}
\algnewcommand\InsideTEE{\Statex \Commentx{Inside TEE.}}
\algnewcommand\OutsideTEE{\Statex \Commentx{Outside TEE.}}
\algnewcommand\EmptyLine{\Statex \vspace{-0.5em}}

\algrenewcommand\algorithmicindent{1.0em}%

\newcommand{\Indent}{\hspace{\algorithmicindent}}

\begin{document}

\CopyrightYear{2016}
\setcopyright{acmauthor}
\isbn{978-1-4503-4670-2/16/12}
\doi{http://dx.doi.org/10.1145/3007788.3007790}

\conferenceinfo{SysTEX '16}{December 12--16, 2016, Trento, Italy}

\title{Proof of Luck: an Efficient \\ Blockchain Consensus Protocol}

\numberofauthors{3}

\author{
\alignauthor
{Mitar Milutinovic}\\
\affaddr{UC Berkeley}\\
\email{mitar@cs.berkeley.edu}
\alignauthor
{Warren He}\\
\affaddr{UC Berkeley}\\
\email{-w@berkeley.edu}
\and
\alignauthor
{Howard Wu}\\
\affaddr{UC Berkeley}\\
\email{howardwu@berkeley.edu}
\alignauthor
{Maxinder Kanwal}\\
\affaddr{UC Berkeley}\\
\email{mkanwal@berkeley.edu}
}

\maketitle
\begin{abstract}

In the paper, we present designs for multiple blockchain
consensus primitives and a novel blockchain system,
all based on the use of trusted execution environments (TEEs), such as
Intel SGX-enabled CPUs.
First, we show how using TEEs for existing proof of work schemes
can make mining equitably distributed by preventing the use of ASICs.
Next, we extend the design with proof of time and proof of ownership
consensus primitives to make mining energy- and time-efficient.
Further improving on these designs, we present a blockchain using a
\emph{proof of luck} consensus protocol.
Our proof of luck blockchain uses a TEE platform's random number
generation to choose a consensus leader, which offers
low-latency transaction validation, deterministic confirmation time,
negligible energy consumption, and equitably distributed mining.
Lastly, we discuss a potential protection against up
to a constant number of compromised TEEs.

\end{abstract}





\section{Introduction}
\label{sec:intro}
Bitcoin~\cite{nakamoto2008bitcoin}, a widely-used blockchain system,
as well as other popular cryptocurrencies~\cite{wood2014ethereum},
have demonstrated that it is practical to use a blockchain as a
distributed ledger.
Maintained by a peer-to-peer network of mutually distrusting participants,
these systems use proof of work~\cite{back2002hashcash} to
solve the key challenge of reaching consensus among participants.

While proof of work is robust against misbehaving and malicious
participants, these algorithms require participants to dedicate
computation time, energy, and silicon towards contrived ``work.''
Moreover, to reduce the number of intermediary forks,
which decrease the effective power of the network,
Bitcoin's proof of work is designed to produce a new block on average every 10 minutes.
It is recommended to wait for 6 blocks before accepting a transaction,~\cite{doublespend}
which makes it impractical for many applications (e.g., point of sale transactions).

To address these shortcomings, we can use
modern trusted execution environments (TEEs) to build new consensus primitives
for use in decentralized electronic currency designs.
The capabilities of TEEs can enforce correct processing of
critical operations and can also limit the effect of Sybils running
under single units of hardware.

In this paper, we propose a novel consensus algorithm,
\emph{proof of luck}, along with a proof-of-concept blockchain design using it,
which achieves low-latency transaction validation while using minimal
energy and computing power under rational attackers and
benign participants. Our design utilizes the capabilities
of Intel's SGX platform, although any TEE platform with similar features will work.
Our design scales to large numbers of participants and sidesteps
the issue of ASICs in mining, allowing
consumer-grade hardware to participate equally in the network.

Our contributions are as follows:
\begin{itemize}
\item We present three basic consensus primitives that use a TEE---proof of work, proof of time, and
proof of ownership. These are energy efficient
drop-in replacements for existing consensus primitives.
\item We present a novel fourth consensus primitive---proof of
luck---and a blockchain design that uses it to achieve low-latency
transaction validation, deterministic confirmation time, negligible
energy consumption, and equitably distributed mining.
\item We discuss a potential protection against an attacker with a
small number of compromised TEEs.
\end{itemize}

Our paper is structured as follows: we
present related work in Section~\ref{sec:relatedwork};
formally define the problem we address in Section~\ref{sec:probdef};
present three drop-in TEE-enabled consensus primitives in Section~\ref{sec:buildingblocks};
present our proof of luck blockchain in Section~\ref{sec:proofofluck};
summarize our analysis in Section~\ref{sec:analysis};
describe future directions in handling TEE compromise in Section~\ref{sec:compromises};
and finally conclude in Section~\ref{sec:conclusion}.

\section{Related Work}
\label{sec:relatedwork}

To address the energy consumption of Bitcoin~\cite{o2013bitcoin}, various alternative
consensus mechanisms have been proposed, e.g., proof of
stake~\cite{king2012ppcoin} and proof of burn~\cite{slimcoin}. It remains
unclear whether these suitably maintain the security properties and incentives
of current proof of work schemes~\cite{poelstra2014distributed}.
Alternative approaches include replacing ``useless'' proof of work puzzles
with meaningful problems~\cite{gridcoin, king2013primecoin},
making energy consumption less wasteful, but still requiring
such computing resources to be available, which limits
mobile and IoT devices from participating.

When there is a known set of participants, the practical Byzantine fault
tolerance algorithm~\cite{castro1999practical}
allows a system to commit transactions and reach consensus within a few
network communications in the presence of up to
one-third of the participants exhibiting arbitrary, malicious behavior.
However, this algorithm
does not scale well in the number of participants, relies critically on
network timing assumptions, and only guarantees liveness when all participants
behave as expected.
While several improvements have been
proposed~\cite{tendermint, stellar, honeybadger}, these still require maintaining
a list of participants or trust relations between subgroups of participants.

To improve transaction confirmation times in Bitcoin, several approaches
have been proposed, e.g., using uncles~\cite{sompolinsky2013accelerating}
and dividing blocks into micro and macro blocks~\cite{eyal2015bitcoin}.
These approaches are orthogonal to our proposed advancements
and can be used to further improve our blockchain system.

Intel is concurrently and independently developing an SGX-based distributed ledger
as part of the Sawtooth Lake project~\cite{sawtooth}.
While it simulates Bitcoin mining in an energy-efficient manner similar
to proof of time, our work provides
additional consensus schemes and benefits, most notably proof of
luck, addresses the issue of possible compromised TEEs, and does not
require estimation of the number of participants.

\section{Problem Definition}
\label{sec:probdef}
The problem of consensus is that the distributed system must agree on
a single shared state.
Current designs of blockchains are slow, using significant time and energy
as part of the consensus mechanism.
Validation prevents arbitrary changes to state,
but still allows the blockchain to ``fork'' into two or more valid
continuations.
A participant may be incentivized to prefer their own state
over other participants' states, e.g. to gain a block mining reward,
or perform a double-spend attack~\cite{karame}.
The system must be able to determine the state of the
participants, uncontrollable by a minority of colluding, malicious participants.

Our goal is to design a consensus algorithm with:

\begin{enumerate}
\item Quick, deterministic transaction confirmations.
\item Energy and network communication efficient protocol.
\item Resistant to custom hardware not commonly available.
\item An attacker controlling under a threshold of
  TEEs cannot control the blockchain.
\item An attacker cannot control the blockchain without controlling
  a majority of CPUs and without breaking the TEE platform.
\item No requirement for a synchronized clock between participants.
\end{enumerate}

We assume the following about the setting of the problem:

\begin{enumerate}
\item The participants have CPUs that implement a suitable TEE, such as Intel SGX.
\item TEE programs can produce unbiased uniform random
  numbers, which an attacker cannot influence. (In Intel SGX, the
  \texttt{RDRAND} instruction is available.)
\item TEE programs can detect concurrent invocations. (In Intel SGX,
  this can currently be done by attempting to create the
  maximum allowed number of monotonic counters for each enclave identity, 256 at the time of writing.)
\end{enumerate}

\subsection{Principals}
\label{sec:principals}

\emph{Participants}, the main principals in our protocol,
are required to use a TEE, and
perform routines to maintain the blockchain and
help others read and write to the blockchain. The
\emph{trusted platform vendor} controls the correct
execution of the algorithm inside each participant's TEE.
\emph{Clients} rely on the blockchain and
communicate with participants to read and write to the blockchain.
As clients do not authenticate participants, they must
protect the content of the blockchain end-to-end, i.e., by
signing any data to be added.

\subsection{Threat Model}
\label{sec:threatmodel}

We analyze our protocol under attack from an adversary controlling
some fraction (less than half) of the participants' machines.
We assume that the adversary cannot break any cryptographic primitives
with non-negligible probability.
It follows that, while the adversary can run the protocol in the TEE
for each controlled machine, they cannot generate a valid attestation
proof if they deviate from the protocol.

In Section~\ref{sec:compromises}, we consider an adversary that can generate
arbitrary valid attestation quotes on a limited number of CPUs.
We base our economic security model of TEE platforms on the fact that
compromising many TEEs is prohibitively expensive.

Furthermore, the adversary can read and send its own network messages,
but cannot modify messages sent by honest participants or cause a network split.
Messages in our protocol do not identify the sender, although some
implementations may choose to do so (e.g., using TCP/IP); in this
case, it is assumed that the adversary can spoof the sender's
information in its own messages and tamper with the sender's information
from honest participants.

Attacks that compromise an entire trusted platform vendor are out of this paper's scope,
as such an attack would have consequences much broader than our blockchain.

\subsection{SGX Background}
\label{sec:sgx}

Our work is based on Intel's SGX platform~\cite{sgx}.
In addition to the isolation and remote attestation features of a
typical TEE,
we utilize the SGX platform's
trusted services that provide relative timestamps and monotonic counters.
SGX uses Enhanced privacy ID (EPID)~\cite{brickell2007enhanced} signatures,
allowing anonymous and pseudonymous attestation.
When initializing a TEE, the platform records a digest
of the code and data (called the \emph{measurement})
which serves as the cryptographic identity of the TEE.
In the remote attestation protocol, the TEE certifies a computation
result with a signed \emph{report} and remotely verifiable
\emph{quote}, which serves as the proof that the results arose from appropriate computations on proper hardware.

\section{Building Blocks}
\label{sec:buildingblocks}

Existing blockchains and other systems utilizing proof of work can
benefit from using these TEE-enabled consensus building blocks directly.
In this section, we
progress through a series of designs that
inform and motivate our proof of luck consensus
protocol, which we present in Section~\ref{sec:proofofluck}.

\subsection{Proof of Work}

One major discussion in the cryptocurrency community is whether
proof of work algorithms should be ASIC-resistant, preventing
custom ASIC hardware from speeding up the computations.
Using a TEE-enabled proof of work sidesteps this issue, as only
supported platforms can be used for mining (e.g., Intel SGX-enabled CPUs).
As a result, mining power is decentralized, mining rewards are equity distributed
among participants, and mining pools bring no advantages.

\begin{algorithm}[h]
  \caption{TEE-enabled proof of work \textbf{(inside TEE)}}
  \begin{algorithmic}[1]
    \Function{PoW}{$\mathit{nonce}$, $\mathit{difficulty}$}
      \State $\mathit{result} \gets \Call{originalPoW}{\mathit{nonce}, \mathit{difficulty}}$
      \State \Assert \Call{originalPoWSuccess}{$\mathit{result}}$
      \State \Return \Call{tee.attestation}{$\langle \mathit{nonce}, \mathit{difficulty} \rangle$, $\textbf{null}$}
    \EndFunction
  \end{algorithmic}
  \label{alg:proofofwork}
\end{algorithm}

We propose Algorithm~\ref{alg:proofofwork}, which wraps any
existing proof of work algorithm into a TEE-enabled one.
Inside a TEE, we call the \textsc{originalPoW} algorithm
with a given $\mathit{nonce}$ and $\mathit{difficulty}$.
In Bitcoin, $\mathit{nonce}$ is the newly-mined block header,
$\mathit{difficulty}$ is the target hash,
and \textsc{originalPoWSuccess} checks that \textsc{originalPoW} succeeded.
\textsc{tee.attestation} returns a $\mathit{proof}$ attesting that the
algorithm ran inside the TEE, and that the code was unmodified (based on the
measurement).
The second argument in our pseudocode, \textbf{null} here, calls for
an anonymous \emph{random base} EPID signature.

\subsection{Proof of Time}
\label{sec:proofoftime}

Proof of work effectively enforces that a sufficient amount of time
has passed by requiring the participant to do work; a TEE can
enforce this directly by requiring the participant to wait for the desired time,
saving CPU cycles and energy for other meaningful work.
We now propose Algorithm~\ref{alg:proofoftime}.

\begin{algorithm}[h]
  \caption{Proof of time \textbf{(inside TEE)}}
  \begin{algorithmic}[1]
    \State $\mathit{counter} \gets \Call{incrementMonotonicCounter}$
    \EmptyLine
    \Function{PoT}{$\mathit{nonce}$, $\mathit{duration}$}
      \State $\Call{sleep}{\mathit{duration}}$
      \State \Assert $\mathit{counter} = \Call{readMonotonicCounter}$
      \State \Return \Call{tee.attestation}{$\langle \mathit{nonce}, \mathit{duration} \rangle$, \textbf{null}}
    \EndFunction
  \end{algorithmic}
  \label{alg:proofoftime}
\end{algorithm}

TEE platforms, like Intel SGX, provide relative timestamps to the TEE.
The \textsc{sleep} function can busy-wait for $\mathit{duration}$,
or it can yield control from the TEE for other outside processes, returning after
$\mathit{duration}$ and verifying that the platform's relative timestamp
has increased sufficiently.

A malicious participant may try to game proof of time by running multiple instances
in parallel on the same CPU.
By maintaining a monotonic counter, provided as a TEE platform service, we can prevent
such malicious behavior by incrementing the counter (\textsc{incrementMonotonicCounter}) every time the TEE starts.
After control is transferred out of the TEE and back, we check
(\textsc{readMonotonicCounter}) that the counter has not changed.

\subsection{Proof of Ownership}
\label{sec:proofofownership}

Proof of work can be seen as a countermeasure against
Sybils, where an attacker acts as multiple
participants---such an attacker would have to do many times the amount
of work.
When participants are required to use a TEE, instead of consuming resources,
one can make it expensive to maintain such virtual participants by
physically limiting each participant to owning a unique CPU.
In SGX, we can operate the EPID signature (part of the remote
attestation protocol) in \emph{name base} mode, which produces a
``pseudonym'' that reveals
if multiple proofs came from the same CPU.
This works even if a malicious user resets the owner epoch register on their CPU.

\begin{algorithm}[h]
  \caption{Proof of ownership \textbf{(inside TEE)}}
  \begin{algorithmic}[1]
    \Function{PoO}{$\mathit{nonce}$}
      \State \Return \Call{tee.attestation}{$\langle \mathit{nonce} \rangle$, $\mathit{nonce}$}
    \EndFunction
  \end{algorithmic}
  \label{alg:proofofownership}
\end{algorithm}

In Algorithm~\ref{alg:proofofownership},
we generate a $\mathit{proof}$ using $\mathit{nonce}$ (a block header)
as the name base (second argument on Line~2).
Proofs generated by a single TEE for the same name base bear the same pseudonym.
A blockchain can then reach consensus by selecting the block that has the most
proofs with unique pseudonyms as the leader.

\section{Proof of Luck}
\label{sec:proofofluck}

Our proof of luck blockchain is built upon the TEE-based proof of luck primitive proposed in Algorithm~\ref{alg:pol}.

\begin{algorithm}[h]
  \caption{Proof of luck primitive \textbf{(inside TEE)}}
  \label{alg:pol}
  \begin{algorithmic}[1]
    \State $\mathit{counter} \gets \Call{incrementMonotonicCounter}{ }$
    \State $\mathit{roundBlock} \gets \textbf{null}$
    \State $\mathit{roundTime} \gets \textbf{null}$
    \EmptyLine
    \Function{PoLRound}{$\mathit{block}$}
      \State $\mathit{roundBlock} \gets block$
      \State $\mathit{roundTime} \gets \Call{tee.getTrustedTime}{ }$
    \EndFunction
    \EmptyLine
    \Function{PoLMine}{$\mathit{header}$, $\mathit{previousBlock}$}
      \Statex \Indent\Commentx{Validating link between $\mathit{header}$ and $\mathit{previousBlock}$.}
      \State \Assert $\mathit{header.parent} = \Call{hash}{\mathit{previousBlock}}$

      \Statex \Indent\Commentx{Validating $\mathit{previousBlock}$ matches $\mathit{roundBlock}$.}
      \State \Assert $\mathit{previousBlock.parent} = \mathit{roundBlock.parent}$

      \Statex \Indent\Commentx{Validating the required time for a round passed.}
      \State $\mathit{now} \gets \Call{tee.getTrustedTime}$
      \State \Assert $\mathit{now} \ge \mathit{roundTime} + \mathit{ROUND\_TIME}$

      \EmptyLine
      \State $\mathit{roundBlock} \gets \textbf{null}$
      \State $\mathit{roundTime} \gets \textbf{null}$
      \State $l \gets \Call{getRandom}$
      \State $\Call{sleep}{f(l)}$

      \EmptyLine
      \Statex \Indent\Commentx{Validating that only one TEE is running.}
      \State $\mathit{newCounter} \gets \Call{readMonotonicCounter}$
      \State \Assert $\mathit{counter} = \mathit{newCounter}$

      \EmptyLine
      \State $\mathit{nonce} \gets \Call{hash}{\mathit{header}}$
      \State \Return \Call{tee.attestation}{$\langle \mathit{nonce}, l \rangle$, $\textbf{null}$}
    \EndFunction
  \end{algorithmic}
\end{algorithm}

The algorithm consist of two functions, \textsc{PoLRound} and \textsc{PoLMine}.
At the start of every round, the participant prepares the TEE to mine on a particular
chain by calling \textsc{PoLRound} and passing the currently known latest
$\mathit{block}$.
After $\mathit{ROUND\_TIME}$ passes, the participant calls \textsc{PoLMine}
to mine a new block.
The participant passes the $\mathit{header}$ of the new block and the
block that it will extend (as $\mathit{previousBlock}$).
$\mathit{previousBlock}$ may be
different from the $\mathit{roundBlock}$ that was passed to \textsc{PoLRound},
but we require that $\mathit{roundBlock}$ and $\mathit{previousBlock}$
have the same $\mathit{parent}$.
This ensures that a participant waits the mandatory $\mathit{ROUND\_TIME}$
between mining blocks, while allowing them to switch to a luckier,
alternative block should they receive one while waiting.
We verify that $\mathit{ROUND\_TIME}$ indeed has passed.

The \textsc{PoLMine} function generates a random value $l \in [0, 1)$ from a
uniform distribution, which is used to determine the winning block from all mined
blocks of participants in this round.
This will tie in with the protocol in Section~\ref{sec:protocol}.

To optimize protocol communication, the algorithm delays releasing
the $\mathit{proof}$ depending on $f(l)$, which prescribes a
shorter delay period for luckier (larger) numbers and a longer wait period for
unluckier numbers.
If a participant receives a luckier block before their own mining
completes, they will not need to broadcast their own block.

We use a monotonic counter as in Section~\ref{sec:proofoftime} 
to prohibit concurrent invocations of the TEE.

\subsection{Protocol}
\label{sec:protocol}

Every participant in the protocol receives transactions from clients and other participants
and maintains a current blockchain representing their view of the system state.
In every round, participants execute an algorithm to commit
pending transactions into a new block, containing a proof of luck
generated from inside a TEE, and extending their current chain.
Participants broadcast their new chain to other participants should the chain
be luckier than any alternative chain received in the meantime.

\begin{algorithm}[h]
  \caption{Extending a blockchain with a new block}
  \label{alg:commit}
  \begin{algorithmic}[1]
    \Function{commit}{$\mathit{newTransactions}$, $\mathit{chain}$}
      \State $\mathit{previousBlock} \gets \Call{latestBlock}{\mathit{chain}}$
      \State $\mathit{parent} \gets \Call{hash}{\mathit{previousBlock}}$
      \State $\mathit{header} \gets \langle \mathit{parent}, \mathit{newTransactions}\rangle$
      \State $\mathit{proof} \gets \Call{PoLMine}{\mathit{header}, \mathit{previousBlock}}$
      \State $\mathit{newBlock} \gets \langle \mathit{parent}, \mathit{newTransactions}, \mathit{proof} \rangle$
      \State \Return $\Call{append}{\mathit{chain}, \mathit{newBlock}}$
    \EndFunction
  \end{algorithmic}
\end{algorithm}

Participants use \textsc{commit} in
Algorithm~\ref{alg:commit} to return a new chain with a $\mathit{newBlock}$ made 
from the set of $\mathit{newTransactions}$.
The $\mathit{newBlock}$ consists of a hash $\mathit{parent}$ of
the previous block, the data $\mathit{newTransactions}$,
and a proof of luck $\mathit{proof}$.

\begin{algorithm}[h]
  \caption{Computing a luck of a valid blockchain}
  \label{alg:luck}
  \begin{algorithmic}[1]
    \Function{luck}{$\mathit{chain}$}
      \State $\mathit{luck} \gets 0$
      \For{$\mathit{block}$ \textbf{in} $\mathit{chain}$}
        \State $\mathit{luck} \gets \mathit{luck} + \Call{tee.proofData}{\mathit{block.proof}}.l$
      \EndFor
      \State \Return $\mathit{luck}$
    \EndFunction
  \end{algorithmic}
\end{algorithm}

The \textsc{luck} algorithm described in Algorithm~\ref{alg:luck} 
computes a numeric score (luck) of a given blockchain by summing the $l$ values of each block.

Our blockchain prefers the chain with the highest luck, which results in desirable behaviors.
First, when appending a block, the new chain is preferred.
Second, when a network split heals, the larger half's chain is likely to have greater luck.
Third, our analysis in Section~\ref{sec:minority} shows that, with
high probability, a minority attacker's chain will not overtake the
majority's.

\begin{algorithm}[h]
  \caption{Validating a blockchain}
  \label{alg:valid}
  \begin{algorithmic}[1]
    \Function{valid}{$\mathit{chain}$}
      \State $\mathit{previousBlock} \gets \textbf{null}$
      \While{$\mathit{chain} \neq \varepsilon$}
        \State $\mathit{block} \gets \Call{earliestBlock}{\mathit{chain}}$
        \State $\langle \mathit{parent}, \mathit{transactions}, \mathit{proof} \rangle \gets block$
        \If{$\mathit{parent} \neq \Call{hash}{\mathit{previousBlock}}$ $\vee$ 
        	\par \Indent \Indent \textbf{not} $\Call{validTransactions}{\mathit{transactions}}$ $\vee$
        	\par \Indent \Indent \textbf{not} $\Call{tee.validAttestation}{\mathit{proof}}$}
          \State \Return \textbf{false}
        \EndIf
        \State $\langle \mathit{nonce}, l \rangle \gets \Call{tee.proofData}{\mathit{proof}}$
        \If{$\mathit{nonce} \neq \Call{hash}{\langle \mathit{parent}, \mathit{transactions} \rangle}$}
          \State \Return \textbf{false}
        \EndIf
        \State $\mathit{previousBlock} \gets \mathit{block}$
        \State $\mathit{chain} \gets \Call{withoutEarliestBlock}{\mathit{chain}}$
      \EndWhile
      \State \Return \textbf{true}
    \EndFunction
  \end{algorithmic}
\end{algorithm}

The \textsc{valid} algorithm in Algorithm~\ref{alg:valid} traverses
the $\mathit{chain}$ from the earliest (genesis) block
(returned from \textsc{earliestBlock}) to the latest block,
ensuring each block has valid transactions, a valid proof of luck,
and a matching previous block hash.
\textsc{tee.validAttestation} is provided by the TEE platform
to validate attestations,
and \textsc{tee.proofData} exposes data used when creating the $\mathit{proof}$.

\begin{algorithm}[h]
  \caption{Proof of luck blockchain protocol}
  \label{alg:pol-protocol}
  \begin{algorithmic}[1]
    \State $\mathit{currentChain} \gets \varepsilon$
    \State $\mathit{transactions} \gets \varepsilon$
    \State $\mathit{roundBlock} \gets \textbf{null}$

    \EmptyLine
    \Function{newRound}{$\mathit{chain}$}
      \State $\mathit{roundBlock} \gets \Call{latestBlock}{\mathit{chain}}$
      \State $\Call{PoLRound}{\mathit{roundBlock}}$
      \State $\Call{resetCallback}{\mathit{callback}, \mathit{ROUND\_TIME}}$
    \EndFunction

    \EmptyLine
    \StartOn{$\mathit{transaction}$ $\textbf{from}$ $\textsc{network}$}
      \If{$\mathit{transaction} \notin \mathit{transactions}$}
        \State $\mathit{transactions} \gets \Call{insert}{\mathit{transactions}, \mathit{transaction}}$
        \State \Call{network.broadcast}{$\mathit{transaction}$}
      \EndIf
    \EndOn

    \EmptyLine
    \StartOn{$\mathit{chain}$ $\textbf{from}$ $\textsc{network}$}
      \If{$\Call{valid}{\mathit{chain}} \wedge \Call{luck}{\mathit{chain}}$ $>$ \par \Indent $\Call{luck}{\mathit{currentChain}}$}
        \State $\mathit{currentChain} \gets \mathit{chain}$
        \If{$\mathit{roundBlock} = \textbf{null}$}
          \State $\Call{newRound}{\mathit{chain}}$
        \Else
          \State $\mathit{latestBlock} \gets \Call{latestBlock}{\mathit{chain}}$
          \If{$\mathit{latestBlock.parent} \ne \mathit{roundBlock.Parent}$}
            \State $\Call{newRound}{chain}$
          \EndIf        
        \EndIf
        \State \Call{network.broadcast}{$\mathit{chain}$}
      \EndIf
    \EndOn

    \EmptyLine
    \StartOn{$\mathit{callback}$}
      \State $\mathit{newTransactions} \gets \mathit{transactions}$
      \State $\mathit{transactions} \gets \varepsilon$
      \State $\mathit{chain} \gets \Call{commit}{\mathit{newTransactions}, \mathit{currentChain}}$
      \State $\Call{network.sendToSelf}{\mathit{chain}}$
    \EndOn
  \end{algorithmic}
\end{algorithm}

We now introduce the proof of luck blockchain protocol in Algorithm~\ref{alg:pol-protocol}. 
Every participant starts with an empty blockchain
$\mathit{currentChain}=\varepsilon$, a set of pending
$\mathit{transactions}=\varepsilon$, and an initial $\mathit{roundBlock}=\textbf{null}$.

After initializing state, participants listen for network messages.
Upon receiving a $\mathit{transaction}$ network message, participants 
add the $\mathit{transaction}$ to their $\mathit{transactions}$ if it has not been already included, and broadcasts it to their peers.
If the message is a new $\mathit{chain}$, they verify that the chain is \textsc{valid}
(Algorithm~\ref{alg:valid}) and has higher \textsc{luck} (Algorithm~\ref{alg:luck}). If so, the participant switches to the new chain and broadcasts it using \textsc{network.broadcast}.

Before broadcasting, the participant may start a new round of mining.
This happens if it is the first round, or if the new chain
has a different $\mathit{parent}$ block from its latest block.
During one round of mining, the participant continues to receive messages about
luckier chains and switches should it have higher luck.
In this instance, the $\mathit{parent}$ will not change.
However, if a participant was part of a network split and just reconciled, the
$\mathit{parent}$ will differ and the participant will have to restart the mining process
on the new chain.

Every time a new round is started (by calling \textsc{newRound}), we call
\textsc{PoLRound} to bind mining to the new chain, clear any pending callbacks,
and schedule a new $\mathit{callback}$ to start after $\mathit{ROUND\_TIME}$.
During normal operation, this means that every participant mines a new block
approximately once every $\mathit{ROUND\_TIME}$ interval.
Participants do not have to have a synchronized clock and their rounds do not
have to be synchronized, but the protocol tends to synchronize rounds.

Inside a $\mathit{callback}$, pending transactions are appended to a new
$\mathit{chain}$ using \textsc{commit} (Algorithm~\ref{alg:commit}) and
the new chain is sent back for processing using \textsc{network.sendToSelf}.

Because the proof of luck TEE releases the proof for luckier (larger)
numbers first, participants will receive new chains
with numbers luckier than their own before receiving their own from
the TEE.
If the participant receives a chain from its TEE that is worse
than the one that they have already received, they do not broadcast its chain
this round.

\section{Analysis}
\label{sec:analysis}

We analyze the behavior of the proof of luck blockchain protocol.

\subsection{Persistence against Minority Attacker}
\label{sec:minority}
We show that several blocks after a fork, it is exponentially
improbable for a minority attacker to produce a chain that is
preferred to a majority of honest participants.

Let $M$ = majority population size, $m$ = minority population size.
For block $t$, we have that the population luck is distributed according to the maximum of uniform random variables:
\begin{align*}
l_M(t) &\stackrel{iid}{\sim} \max\{\{\textrm{Uniform}(0,1)\}^M\}\\
l_m(t) &\stackrel{iid}{\sim} \max\{\{\textrm{Uniform}(0,1)\}^m\}
\end{align*}

After $h$ blocks from a fork, we define the relative total luck:
\begin{align*}
L^{(h)} := \sum_{t=1}^h l_M(t) - l_m(t)
\end{align*}

Using a Chernoff bound and independence assumptions, we now show that the probability of the event that a minority wins is exponentially small in $h$:
\begin{align*}
Pr\left(L^{(h)} \leq 0\right) &\leq \min_{s > 0} \mathbb{E} \left[e^{- s L^{(h)}}\right]\\
&= \min_{s > 0} \prod_{t=1}^{h} \mathbb{E} \left[e^{- s l_M(t)}\right] \mathbb{E} \left[e^{s l_m(t)}\right]\\
&= \min_{s > 0} \left( \mathbb{E} \left[e^{- s l_M(t)}\right] \mathbb{E} \left[e^{s l_m(t)}\right] \right)^h
\end{align*}

Because $M > m$, there exists an $s > 0$ such that the product of the inner expectations is less than $1$.

Hence we have that the probability of the event that the total minority luck
exceeds the total majority luck decreases exponentially in the number of blocks, $h$, after a fork.

\subsection{Proportional Control of Blocks}
Consider a group of participants $A\subseteq G$, from all participants $G$.
Among honest participants that append to the longest chain, the new
chain with the largest luck value $l$ in the newly added block is preferred.
At each round, every participant has an equal probability of generating
the largest random number, because the participants sample
independently from identical distributions.
Thus over time, the expected number of blocks mined by group $A$ is
proportional to the number of participants in $A$.

\subsection{Round Time and Confirmation Time}

We propose a $\mathit{ROUND\_TIME}$ of 15 seconds, which creates
block confirmation times slightly larger than 15 seconds, comparable
with Ethereum and much faster than Bitcoin's 10 minutes.
We have chosen this value based on an evaluation of information in the
Bitcoin network~\cite{decker2013information}, where median block propagation time has been observed
to be around $6.5$ seconds.
The initial selection of the winning block with maximal luck value $l$ can be
implemented without transmitting whole blocks, only block headers (i.e., without the full transaction data), which should take
just one round-trip based on Bitcoin's transaction propagation time,
and only after a winner is determined should the whole block be propagated.

\section{Compromised TEE}
\label{sec:compromises}

Our consensus protocol assumes that the security requirements of the
TEE are met for all participants.
Although it would be expensive to violate these requirements, a
motivated attacker may compromise a limited number of TEEs.
We discuss a mitigation extension to our protocol.

\subsection{Luckiest $m$}

To strengthen our design against high-cost attacks on individual TEEs,
we discuss a possible extension to our protocol, which
constructs a blockchain consisting of \textit{super-blocks}:
super-blocks are made by merging $m$ normal blocks and their proofs of luck.
Participants continue to create individual blocks based on pending transactions and
attach a proof of luck with a luck value $l$ to the block.
However, participants now select the $m$ luckiest blocks ($m$ blocks with
highest $l$) and merge, in a deterministic way, those $m$ blocks
into a super-block.

Each super-block has $m$ proofs of luck, with values $l_{1},...,l_{m}$, where
$l_{1}>...>l_{m}$ (i.e., $l_{1}$ is the luckiest).
Under this extension, the luck of the super-block as a whole is $l_m$,
the least luckiest value. This ensures that even if $m-1$ CPUs 
have been compromised, they still cannot fully control the $l$ value of the super-block.
Future work may analyze the persistence propery of this extension.

This approach is similar to our proof of ownership primitive from
Section~\ref{sec:proofofownership}. However, instead of recording
proofs from all participants, we record only the luckiest $m$,
making this approach more scalable.
In our SGX-based prototype, we
ensure that the $m$ proofs come from different CPUs
by having the
participants generate linkable quotes during remote attestation, using
the parent block's hash as a
name base in the EPID signature scheme.
EPID signatures produce a pseudonym based on the name base and the
CPU's private key, and proofs with the same pseudonym would reveal
that they came from the same physical CPU.
Thus, even if a CPU is compromised, it is still limited to supplying
one proof of luck per super-block.

\subsection{Merging}

After the honest participants produce their own blocks, they broadcast
their block to other participants that chose the same
blockchain to extend (they may choose different blockchains due to
network latency).
Participants then merge the $m$ luckiest blocks into a super-block.
Blocks from honest participants have nearly the same transactions, so
a super-block can often be compressed efficiently.
Having participants merge blocks into super-blocks is similar to
Ethereum's concept of block uncles, in which a block may have multiple
parents, and which improves the throughput and liveness of the
protocol.

\section{Conclusion}
\label{sec:conclusion}

We have proposed three TEE-enabled building block designs 
for systems using existing proof of work schemes: TEE-enabled
proof of work, proof of time, and proof of ownership.
We have combined ideas from these primitives into a new blockchain system 
based on the proof of luck consensus protocol.
Our analysis shows that our blockchain
ensures liveness and persistence, while providing 
energy efficient mining, low-latency transaction validation
with deterministic confirmation time, and decentralized mining
power.
Lastly, we discuss a potential protection against a constant number of
compromised TEEs participating in the blockchain.

\section{Acknowledgements}
\label{sec:acknowledgements}

We thank Dawn Song for helpful suggestions.
This work was supported by IC3, NSF (TWC-1518899), DARPA (N66001-15-C-4066).
Any opinions, findings, and conclusions or recommendations expressed in this material are those of the author(s)
and do not necessarily reflect the views of NSF and/or DARPA.

\bibliographystyle{abbrv}
\bibliography{refs}

\end{document}